%% file: project.tex
\documentclass[journal,twocolumn,twoside,10pt]{IEEEtran}
\input{header}

\begin{document}
%
\title{A Non-iterative Method for (Re)Construction of Phase from STFT Magnitude}
%
%
%

\author{Zden\v{e}k Pr\r{u}\v{s}a, 
        Peter Balazs,~\IEEEmembership{Senior Member,~IEEE,} and
        Peter L. S\o ndergaard
\thanks{Manuscript received April 19, 2005; revised August 26, 2015.}
\thanks{Z. Pr\r{u}\v{s}a* and P. Balazs are with the
Acoustics Research Institute, Austrian Academy of Sciences,
Wohllebengasse 12--14, 1040 Vienna, Austria, email:
\texttt{zdenek.prusa@oeaw.ac.at} (corresponding address),
\texttt{peter.balazs@oeaw.ac.at}}
\thanks{P. L. S\o ndergaard is with Oticon A/S, Kongebakken 9, 2765 Sm{\o}rum, 
Denmark, email: \texttt{peter@sonderport.dk}}
\thanks{This work was supported by the Austrian Science Fund
(FWF) START-project FLAME (``Frames and Linear Operators for Acoustical
Modeling and Parameter Estimation''; Y~551-N13).}
}

%
%

\markboth{IEEE/ACM Transactions on Audio, Speech, and Language Processing}%
{Pr\r{u}\v{s}a \MakeLowercase{\textit{et al.}}: A non-iterative Method for
(Re)Construction of Phase from STFT Magnitude}
%



\maketitle

\begin{abstract}
    \input{abstract}
\end{abstract}

\IEEEpeerreviewmaketitle

\input{main}

\bibliographystyle{IEEEtran}
\bibliography{project}

\end{document}

%% file: header.tex
\usepackage{graphicx}
\usepackage{enumerate}
\usepackage{amsmath}
\usepackage{amsfonts}
\usepackage{amsthm}
\usepackage{commath}
\usepackage[ruled]{algorithm2e}
\let\oldnl\nl
\newcommand{\nonl}{\renewcommand{\nl}{\let\nl\oldnl}}
\usepackage{amscd}
\usepackage{url}
\usepackage{subfig}
\usepackage{color}

\theoremstyle{definition}

\theoremstyle{plain}

\newtheorem{proposition}{Proposition}

\theoremstyle{remark}

\definecolor{violet}{rgb}{0.33,0.10,0.54} 
\newcommand{\xxl}[1]{#1}

\DeclareMathOperator*{\argmax}{arg\,max}

\newcommand{\stime}{t}
\newcommand{\stimearg}{\tau}
\newcommand{\sfreq}{\omega}

\newcommand{\dstimearg}{\ensuremath{\, \mathrm{d}\stimearg}}

\newcommand{\modop}{\ensuremath{\mathcal{E}}}
\newcommand{\tranop}{\ensuremath{\mathcal{T}}}

\newcommand{\modoparg}[1]{\ensuremath{\modop_{#1}}}
\newcommand{\tranoparg}[1]{\ensuremath{\tranop_{#1}}}

\newcommand{\modopsfreq}{\modoparg{\sfreq}}
\newcommand{\tranopstime}{\tranoparg{\stime}}

\newcommand{\stftphase}[2]{\ensuremath{\Phi_{#1}^{#2}}}
\newcommand{\stftmodulus}[2]{\ensuremath{M_{#1}^{#2}}}

\newcommand{\stftphasefg}{\stftphase{g}{f}}
\newcommand{\stftmodulusfg}{\stftmodulus{g}{f}}

\newcommand{\me}{\mathrm{e}}
\newcommand{\mi}{\mathrm{i}}



%% file: abstract.tex
A non-iterative method for the construction of the Short-Time Fourier 
Transform (STFT) phase from the magnitude is presented.
The method is based on the direct relationship
between the partial derivatives of the phase and the logarithm of the magnitude
of the un-sampled STFT with respect to the Gaussian window.
Although the theory holds in the continuous setting only,
the experiments show that the algorithm performs well
even in the discretized setting (Discrete Gabor transform) with low redundancy
using the sampled Gaussian window, the truncated Gaussian window and even other
compactly supported windows like the Hann window.

Due to the non-iterative nature,
the algorithm is very fast and it is suitable for long audio signals.
Moreover, solutions of iterative phase reconstruction algorithms
can be improved considerably by initializing them
with the phase estimate provided by the present algorithm.

We present an extensive comparison with the
state-of-the-art algorithms in a reproducible manner.

%% file: main.tex
\section{Introduction}
The phase retrieval problem has been actively investigated for decades.
It was first formulated for the Fourier transform and later
for generic linear systems.

%

In this paper, we consider a particular case of the phase retrieval problem;
the reconstruction from the magnitude of the Gabor transform coefficients
obtained by sampling the STFT magnitude at
discrete time and frequency points \cite{chabible}.
The need for an effective way of the phase (re)construction arises in
audio processing applications such as 
source separation and denoising \cite{guse10,stda12}, 
time-stretching/pitch shifting \cite{lado99},
channel mixing \cite{gnsp08},
and missing data inpainting \cite{smbima11}.

The problem has already been addressed by many authors.
Among the iterative algorithms,
the most widespread and influential is the Griffin-Lim algorithm (GLA)
\cite{griflim84} which inspired several extensions \cite{zhbewy07,leroux10,pebaso13}.
See \cite{stda11} for a detailed overview of the algorithms based on the idea of
Griffin and Lim.
A different approach was taken in \cite{desomada15}, where the authors proposed to
express the problem as an unconstrained
optimization problem and to solve it using the limited memory
Broyden-Flatcher-Goldfarb-Shanno algorithm.
It is again an iterative algorithm and the computational cost of a single
iteration is comparable to that of GLA.

Other approaches are based on reformulating the task as a convex problem
\cite{wadama15,caelstvo15,susm12,ba10}. 
The dimension of the problem however squares, 
which makes it unsuitable for long audio signals
which typically consist of tens of thousands of samples per second.

The approach from the recently published work \cite{elsimibaco15} builds upon
the assumption that the signal is sparse in the original domain, which is not
realistic in the context of the audio processing applications mentioned above.

An interesting approach was presented in \cite{boez11}. It is based on
solving non-linear system of equations for each time frame. The authors proposed
to use iterative solver and initialize it with samples obtained from 
previous frames.
The algorithm is, however, designed to work exclusively with a rectangular 
window.



The common problem of the iterative state-of-the-art algorithms is that
they require many relatively
expensive iterations in order to produce acceptable results.
Recently, a non-iterative algorithm was proposed in \cite{be15}.
It is based on the notion of \emph{phase consistency}
used in the phase vocoder \cite{lado99}.
Although the algorithm is simple, fast and it is directly suitable for the real-time
setting, it relies on the fact that the signal consists of slowly varying sinusoidal
components and fails for transients and broadband components in general.
A similar algorithm was introduced in \cite{mabada15}, which, in addition,
tries to treat impulse-like components separately.

In this paper, we propose another
non-iterative algorithm (Phase Gradient Heap Integration -- PGHI). 
The theory behind PGHI has been known at
least since 1979. Indeed, it is based on the relationship between gradients of the
Gaussian window-based STFT phase and log-magnitude
published already in \cite{po79} and on the \emph{gradient theorem}.
More precisely, the phase gradient can be expressed using the
STFT magnitude and the gradient theorem gives a prescription how to integrate the 
phase gradient field to recover the phase up to a global phase shift
(or up to sign ambiguity in case of real signals).
To our knowledge, no such algorithm has ever been published yet.
Curiously enough, in \cite{no10} it was even explicitly discouraged to use such
algebraic results for practical purposes.

The aforementioned algorithms \cite{be15} and \cite{mabada15}
are in fact close to the PGHI algorithm since they
basically perform a crude integration of the estimate of \emph{instantaneous frequency}
and of the \emph{local group delay} in case of \cite{mabada15},
which are components of the STFT phase gradient.

In the spirit of reproducible research,
the implementation of the algorithms,
audio examples, color version of the figures as well as
scripts reproducing experiments from this manuscript are freely
available at \url{http://ltfat.github.io/notes/040}.
The code depends on our
Matlab/GNU Octave\cite{octave} packages LTFAT \cite{ltfatnote015,ltfatnote030}
(version 2.1.2 or above) and PHASERET (version 0.1.0 or above).
Both toolboxes can be obtained freely 
from \url{http://ltfat.github.io} and \url{http://ltfat.github.io/phaseret},
respectively.



The paper is organized as follows. Section~\ref{sec:gab} summarizes the necessary
theory of the STFT and the Gabor analysis, Section~\ref{sec:theory} presents the theory
behind the proposed algorithm, Section~\ref{sec:algorithm} contains a detailed description
of the numerical algorithm. Finally, in Section~\ref{sec:exp}
we present an extensive evaluation of the proposed algorithm and
comparison with the iterative and non-iterative state-of-the-art algorithms
using the Gaussian window, the truncated Gaussian window, the Hann and 
the Hamming windows.

\section{Gabor analysis}\label{sec:gab}
The \emph{short-time Fourier transform}
of a function $f \in L^2(\mathbb{R})$ with respect to a window
$g \in L^2 (\mathbb{R})$ can be defined as%
\footnote{In the literature, two other
    STFT phase conventions can be found. The present one is 
most common in the engineering community.}
\begin{align} \label{eq:shorttime}
    (\mathcal{V}_g f) (\sfreq, \stime)
    &= \int_{\mathbb{R}}\! f(\stimearg)
    \overline{g(\stimearg-\stime)} \me^{-\mi 2\pi \sfreq (\stimearg - \stime) } \,
    \mathrm{d}\stimearg,
    \ \ \sfreq, \stime \in\mathbb{R}\\
    &=: \stftmodulusfg(\sfreq,\stime) \cdot \me^{\mi \stftphasefg(\sfreq,\stime)},
\end{align}
\xxl{where we have separated the amplitude and phase component.}
Using the modulation $\left(\modoparg{\sfreq} f\right) (\stimearg) := \me^{\mi 2 \pi
\sfreq \stimearg} \cdot f(\stimearg)$ and translation $\left(\tranoparg{\stime} f\right) (\stimearg) :=
f (\stimearg - \stime)$ we get the alternative representation
$\left(\mathcal{V}_g f\right) (\sfreq, \stime) = \left< f ,   \tranoparg{\stime}\modoparg{\sfreq} g \right>$.

The (complex) logarithm of the STFT can be written as 
\begin{equation}
    \log(\mathcal{V}_g f) (\sfreq, \stime) = \log \stftmodulusfg(\sfreq,\stime) + \mi
    \stftphasefg(\sfreq,\stime).
    \label{eq:shorttimelog}
\end{equation}

The \emph{Gaussian function} is a particularly suitable window function as
it possesses optimal time-frequency properties and
it allows an algebraic treatment of the equations.
It is defined by the following formula
\begin{align}
    \varphi_\lambda (t) = \left(\frac{\lambda}{2}\right)^{-\frac{1}{4}} \me^{-\pi
    \frac{t^2}{\lambda}}=\left(D_{\sqrt{\lambda}}\varphi_1\right)(t),
\end{align}
where $\lambda \in \mathbb{R}^+$ denotes the ``width'' or the time-frequency ratio
of the Gaussian window and $D_\alpha$ is a dilation operator such that
$(D_\alpha f)(t) = \frac{1}{\sqrt{|\alpha|}} f(t/\alpha)$, $\alpha\neq0$.
We will use the shortened notation $\varphi(t) = \varphi_1(t)$ in the following text.

\subsection{Discrete Gabor Transform}
\label{sec:dgt}
We define the Discrete Gabor Transform (DGT) of a signal $f\in\mathbb{C}^L$
with respect to a window $g\in\mathbb{C}^L$ as \cite{stroh1,ltfatnote006,ltfatnote003}
\begin{align}
          c(m,n) &= \sum_{l=0}^{L-1} f(l)\overline{g(l-na)}\me^{-\mi 2 \pi m (l-na)/ M} \label{eq:dgt} \\
           &=: s(m,n) \cdot \me^{\mi\phi(m,n)} \label{eq:dgtpolar}
\end{align}
for $m=0,\dots,M-1$ and $n=0,\dots,N-1$, where $M=L/b$ is the number of frequency channels,
$N=L/a$ number of time shifts, $a$ is the length of the time shift or a hop size in
samples in time and $b$ is a hop size in samples in frequency. The bar denotes complex conjugation and
$(l-na)$ is assumed to be evaluated modulo $L$.  The redundancy of the DGT is defined as $MN/L=M/a$.  
In the matrix notation, we can write
$c = F_{g}^*f$, where $F_{g}^*$ is a $MN\times L$ matrix. (Note that this matrix has a very particular block-structure \cite{xxlfei1}.)

The DGT can be seen as sampling of STFT
(both of the arguments $\sfreq$ and $\stime$ and the involved functions $f$ and $g$ themselves) 
of one period of
$L$-periodic continuous signal $f$ such that
\begin{equation}
    c(m,n) = \left(\mathcal{V}_{g}f\right)(b m, a n) +
    \mathcal{A}(m,n),
\end{equation}
for
$m = 0,\dots,M-1$, $n = 0,\dots,N-1$
where $\mathcal{A}(m,n)$ models both the aliasing
and numerical errors introduced by the sampling.

For real signals $f\in\mathbb{R}^L$, the range of $m$ can be shrunken to the first
$\lfloor M/2 \rfloor + 1$ values as the remaining coefficients are
complex conjugated. Moreover, the coefficients $c(0,\bullet)$ are always real and so are
$c(M/2,\bullet)$ if $M$ is even.

Signal $f$ can be recovered (up to a numerical precision error) using the
following formula
\begin{equation}
    f(l) = \sum_{m=0}^{M-1}\sum_{n=0}^{N-1}c(m,n)\widetilde{g}(l-na)\me^{\mi2\pi m
    \left(l-na\right) /M}
    \label{eq:idgt}
\end{equation}
for $l=0,\dots,L-1$,
where $\widetilde{g}$ is the canonical dual window. 
In the matrix notation,
we can write $f=F_{\widetilde{g}}c$, where $F_{\widetilde{g}}$
is a $L\times MN$ matrix. The canonical dual window can be obtained as
\begin{equation}
    \widetilde{g}=\left(F_gF^*_g\right)^{-1} g.
    \label{eq:dualwin}
\end{equation}
See e.g. \cite{ltfatnote003} for
conditions under which the product $F_gF^*_g$ is (easily) invertible and \cite{ltfatnote011}
for an overview of efficient algorithms for computing \eqref{eq:dgt}, \eqref{eq:idgt} and
\eqref{eq:dualwin}. In particular the block structure can be used for a pre-conditioning approach \cite{xxlfei1}.

The discretized and periodized Gaussian window is given by
\begin{equation}
    \varphi_\lambda^{\text{D}} (l) = \left(\frac{\lambda L}{2}\right)^{-\frac{1}{4}} \sum_{k\in
    \mathbb{Z}} \me^{-\pi\frac{\left(l+kL\right)^2}{\lambda L}},
    \label{eq:pgauss}
\end{equation}
for $l=0,\dots,L-1$. We assume that $L$ and $\lambda$ are chosen such 
that the time aliasing is numerically negligible and 
therefore it is sufficient to sum over $k\in\left\{-1,0\right\}$
in practice.

The width of the Gaussian window at its 
relative height $h \in [0,1]$ is given by
\begin{equation}
    w_h = \sqrt{-\frac{4\log(h)}{\pi}\lambda L}.
    \label{eq:gausssamples}
\end{equation}
The width is given in samples and it can be a non-integer number.
This equation becomes relevant when working with truncated Gaussian window or  
with other non-Gaussian windows.

Note that all windows used in this manuscript are odd symmetric,
such that they have a unique center sample, and they are non-causal such that
they introduce no delay.
Finally, the discrete Fourier transform of such windows is real.


\section{Theory Behind The Algorithm}\label{sec:theory}
The algorithm is based on the direct relationship
between the partial derivatives of the phase and the log-magnitude of the STFT with
respect to the Gaussian window. In this section, we derive such relations.
We include a complete derivation since the relations 
for the STFT as defined in \eqref{eq:shorttime} has not appeared in the literature, 
as far as we know.  

It is known that the Bargmann transform of $f\in L^2(\mathbb{R})$
\begin{align}
    \left(\mathcal{B}f\right) (z) = 2^{\frac{1}{4}} \int_{\mathbb{R}} f(\stimearg)
    \me^{2\pi \stimearg z - \pi \stimearg^2 - \frac{\pi}{2} z^2} \, \mathrm{d}\stimearg, \quad
    z \in \mathbb{C}
\end{align}
is an entire function \cite{complexanalysis78} for all $z\in\mathbb{C}$ and 
that it relates to the STFT defined in \eqref{eq:shorttime} such that 
\begin{align}
    (\mathcal{B}f)(z)=\me^{ \pi \mi \stime \sfreq +\pi \frac{|z|^2}{2}}
    (\mathcal{V}_{\varphi}f)(-\sfreq, \stime),
    \label{eq:stftentire}
\end{align}
assuming $z=\stime+\mi\sfreq$ \cite{chabible}.
The logarithm of the Bargmann transform is an entire function as well (apart from zeros).
The real and imaginary parts of $\log(\mathcal{B}f)(z)$ can be written
as
\begin{align}
    \log(\mathcal{B}f)(\stime + \mi\sfreq)
    &=u(\sfreq,\stime)+\mi v(\sfreq,\stime)\\
    u(\sfreq,\stime)&= \pi(\stime^2 +\sfreq^2)/2 + \log M_{\varphi}^f(-\sfreq, \stime) \\
    v(\sfreq,\stime)&= \pi \stime \sfreq + \Phi_\varphi^f(-\sfreq,\stime)
\end{align}
and using the Cauchy-Riemann equations
\begin{align}
    \pd{u}{\stime}(\sfreq,\stime) &= \pd{v}{\sfreq}(\sfreq,\stime) \\
    \pd{u}{\sfreq}(\sfreq,\stime) &= -\pd{v}{\stime}(\sfreq,\stime)
\end{align}
we can write (substituting $\sfreq'=-\sfreq$) that
\begin{align} \label{eq:origderiv1}
    \pd{}{\sfreq'}\Phi_\varphi^f(\sfreq',\stime) &= - \pd{
    }{\stime}\log M_{\varphi}^f(\sfreq',\stime)  
\\ \label{eq:origderiv2}
\pd{}{\stime}\Phi_\varphi^f(\sfreq',\stime) &=
    \pd{}{\sfreq'}\log M_{\varphi}^f(\sfreq',\stime) + 2\pi\sfreq'.
\end{align}

A little more general relationships can be obtained for windows defined as
$g=\mathcal{O}\varphi_1$
($\mathcal{O}$ being a fixed bounded operator) and Proposition \ref{prop:gencauchy}.

\begin{proposition}
    \label{prop:gencauchy}
    Let $\mathcal{O}, \mathcal{P}$ be bounded operators such that for all
    $(\sfreq,\stime)$
    there exist differentiable, \xxl{strictly monotonic} functions $\eta(\stime)$ and $\xi(\sfreq)$,
    such that $\tranopstime \modopsfreq\mathcal{O} = \mathcal{P}
     \tranoparg{\eta(\stime)}\modoparg{\xi(\sfreq)}$ and
    let $g=\mathcal{O}\varphi_1$. Then
    \begin{align}
        \pd{}{\sfreq} \Phi^f_g(\sfreq,\stime) &=
        -\pd{}{\stime}\log M^f_g(\sfreq,\stime)\cdot\frac{\xi'(\sfreq)}{\eta'(\stime)} \\
        \pd{}{\stime} \Phi^f_g(\sfreq,\stime) &=\pd{}{\sfreq}\log
        M^f_g(\sfreq,\stime)
        \cdot\frac{\eta'(\stime)}{\xi'(\sfreq)}
        + 2\pi \xi(\sfreq)\eta'(\stime).
    \end{align}
    
    \begin{proof}
Consider
\begin{align*}
    \left(\mathcal{V}_{g}f\right)(\sfreq, \stime) & =  \left\langle f, \tranopstime \modopsfreq
    g\right\rangle = \left\langle f,\tranopstime \modopsfreq \mathcal{O} \varphi_1 \right\rangle\\
    & =  \left\langle \mathcal{P}^{*}\hspace{-0.2em}f, \tranoparg{\eta(\stime)} \modoparg{\xi(\sfreq)} \varphi_1 \right\rangle \\
    & =  \left(\mathcal{V}_{\varphi_1}\left(\mathcal{P}^{*}\hspace{-0.2em}f\right)\right)\left(\xi(\sfreq),\eta(\stime)\right)
 \end{align*}
Therefore 
\begin{align*}
    \pd{}{\stime} \Phi^f_g(\sfreq,\stime)  & = \pd{}{\stime}  \left[ \Phi^{ \mathcal{P}^{*}\hspace{-0.2em}f }_{\varphi_1} \left(\xi(\sfreq),\eta(\stime)\right) \right]  \\
                                           & = \left[\pd{}{\eta} \Phi^{ \mathcal{P}^{*}\hspace{-0.2em}f }_{\varphi_1} \left(\xi(\sfreq),\eta(\stime)\right)\right] \cdot \eta'(\stime)
\end{align*}
 and 
\begin{align*}
    \pd{}{\sfreq} \Phi^f_g(\sfreq,\stime)  = \left[\pd{}{\xi} \Phi^{ \mathcal{P}^{*}\hspace{-0.2em}f
    }_{\varphi_1} \left(\xi(\sfreq),\eta(\stime)\right)\right] \cdot \xi'(\sfreq).
\end{align*}
Furthermore
 \begin{align*}
     \pd{}{\stime}\log M^f_g(\sfreq,\stime)  =  \left[\pd{}{\eta}\log M^{
 \mathcal{P}^{*}\hspace{-0.2em}f }_{\varphi_1}
 \left(\xi(\sfreq),\eta(\stime)\right)\right] \cdot \eta'(\stime)
  \end{align*}
  and
  \begin{align*}
      \pd{}{\sfreq}\log M^f_g(\sfreq,\stime)  =  \left[\pd{}{\xi}\log M^{
      \mathcal{P}^{*}\hspace{-0.2em}f }_{\varphi_1} \left(\xi(\sfreq),\eta(\stime)\right)\right] \cdot \xi'(\sfreq).
 \end{align*}
 Combining this with \eqref{eq:origderiv2} we get
 \begin{align*}
 \pd{}{\stime} \Phi^f_g(\sfreq,\stime)  & = \left[ \pd{}{\eta} \Phi^{
 \mathcal{P}^{*}\hspace{-0.2em}f }_{\varphi_1} \left(\xi(\sfreq),\eta(\stime)\right) \right]  \cdot \eta'(\stime) \\
                                        & =  \left[ \pd{}{\xi}\log M^{ \mathcal{P}^{*}\hspace{-0.2em}f }_{\varphi_1}
                                    \left(\xi(\sfreq),\eta(\stime)\right) + 2\pi \xi(\sfreq)\right] \cdot \eta'(\stime) \\
 & =  \pd{}{\sfreq}\log M^f_g(\sfreq,\stime) \cdot \frac{\eta'(\stime)}{\xi'(\sfreq)} +
     2\pi \xi(\sfreq) \eta'(\stime).
 \end{align*}
 The other equality can be shown using the same arguments and \eqref{eq:origderiv1}. 
    \end{proof}
\end{proposition}

Choosing $\mathcal{O}=D_{\sqrt{\lambda}}$, 
$\xi(\sfreq)=\sqrt{\lambda}\sfreq$ and
$\eta(\stime) = \stime/\sqrt{\lambda}$
leads to equations for dilated Gaussian window $\varphi_\lambda$
\begin{align}
    \pd{}{\sfreq}\Phi^{f}_{\varphi_\lambda}(\sfreq,\stime) &=
    -\lambda\pd{}{\stime}\log M^f_{\varphi_\lambda}(\sfreq,\stime)
    \label{eq:phiomega} \\
    \pd{}{\stime}\Phi^f_{\varphi_\lambda}(\sfreq,\stime) &=
    \frac{1}{\lambda}\pd{}{\sfreq}\log M^f_{\varphi_\lambda}(\sfreq,\stime) + 2\pi\sfreq
    \label{eq:phit}.
\end{align}
The relations were already published in \cite{po79,gama06,auchfl12} 
in slightly different forms obtained using different techniques than we use here. 
The equations differ because the authors of the above mentioned papers use different STFT
phase conventions.

It should be noted that the relations for general windows were already studied in
\cite{cmdaaufl97}, they however involve partial derivatives of the logarithm of the
modified Bargmann transform and thus it seems they cannot be exploited directly.
Moreover, the experiments presented in Section~\ref{sec:exp} show that the performance degradation
is not too significant when using windows resembling the Gaussian window.

The STFT phase gradient of a signal $f$ with respect to dilated Gaussian $\varphi_\lambda$
will be further denoted as
\begin{equation}
    \nabla\Phi_{\varphi_\lambda}^f(\sfreq,\stime) = \left[\pd{}{\sfreq}\Phi_{\varphi_\lambda}^f(\sfreq,\stime),
    \pd{}{\stime}\Phi_{\varphi_\lambda}^f(\sfreq,\stime)\right].
\end{equation}
Note that the derivative of the phase has a peculiar pole pattern around zeroes \cite{xxlbayjailsoend11}.
%


\subsection{Gradient Integration and the Phase Shift Phenomenon}
Knowing the phase gradient, one can exploit the gradient theorem 
(see e.g.  \cite{gradtheorem15})
to recover the original phase $\Phi^f_{\varphi_\lambda}(\sfreq,\stime)$ such that
\begin{equation}
    \Phi^f_{\varphi_\lambda}(\sfreq,\stime) - \Phi^f_{\varphi_\lambda}(\sfreq_0,\stime_0)
    = \int_0^1 \!
    \nabla\Phi^f_{\varphi_\lambda}\left(r\left(\stimearg\right)\right)\cdot
    \od{r}{\stimearg}\left(\stimearg\right) \dstimearg,
    \label{eq:lineint}
\end{equation}
where $r(\stimearg)=[r_{\sfreq}(\stimearg),r_{\stime}(\stimearg)]$ is any curve
starting at $(\sfreq_0,\stime_0)$ and ending at $(\sfreq,\stime)$
provided the phase at the initial point $(\sfreq_0,\stime_0)$ is known.
When the phase is unknown completely, we consider
$\Phi^f_{\varphi_\lambda}(\sfreq_0,\stime_0)=0$ and therefore the phase one obtains by
\eqref{eq:lineint} is 
\begin{equation}
    \widetilde{\Phi}^{f}_{\varphi_\lambda}(\sfreq,\stime) = \Phi^f_{\varphi_\lambda}(\sfreq,\stime) + \Phi_0,
\end{equation}
where $\Phi_0=\Phi^f_{\varphi_\lambda}(\sfreq_0,\stime_0)$ is a constant global phase shift.

The global phase shift of the STFT carries over to the global phase shift
of the reconstructed signal trough the linearity of the reconstruction.
One must, however, treat real input signals with
care as the phase shift breaks the complex conjugate relation of the 
positive and negative frequency coefficients. 
This relationship has to be either recovered or enforced because if
one simply takes only the real part of the reconstructed signal
the phase shift causes its amplitude attenuation or even causes the signal 
to vanish in the extreme case.
To explain this phenomenon, consider the following example where we compare the
effect of the phase shift on analytic and on real signals.
We denote the constant phase shift as $\psi_0$ and define an analytic signal
as $x_{\text{an}}(t)=A(t)\me^{\mi\psi(t)}$. The real part including the global phase
shift ($\me^{\mi \psi_0}$) is given as
$\mathcal{R}( x_{\text{an}}(t)\me^{\mi \psi_0} ) = A(t)\cos(\psi(t)+\psi_0)$
which is what one would expect.
Similarly, we define a real signal as
$x(t)=\frac{A(t)}{2}\left(\me^{\mi\psi(t)}+e^{-\mi\psi(t)}\right)$ and the real
part of such signal with the global phase shift $\psi_0$ amounts to
$\mathcal{R}( x(t)e^{\mi \psi_0} ) = A(t)\cos(\psi_0)\cos(\psi(t))$
which causes the signal to vanish when $\psi_0 = \pi/2 + k\pi$, $k\in\mathbb{Z}$.

In theory, the global phase shift of the STFT of a real signal can be compensated
for, leaving only a global signal sign ambiguity.
For real signals, it is clear that the following holds for $\sfreq \neq 0$
\begin{equation}
    \widetilde{\Phi}^{f}_{\varphi_\lambda}(\sfreq,\stime) +
    \widetilde{\Phi}^{f}_{\varphi_\lambda}(-\sfreq,\stime)
    =  2 \Phi_0.
    \label{eq:realconj}
\end{equation}
Due to the phase wrapping, after the compensation,
the phase shift is still ambiguous such that
$\Phi_0=k\pi$ for $k\in\mathbb{Z}$, which causes the aforementioned signal sign
ambiguity.


\section{The Algorithm}\label{sec:algorithm}
In the discrete time setting (recall Section~\ref{sec:dgt}; in particular \eqref{eq:dgt}
and \eqref{eq:dgtpolar})
the STFT phase gradient approximation
$\widehat{\nabla\Phi_{\varphi_\lambda}}(b m,a n) := \nabla\phi(m,n)$
is obtained by numerical differentiation of
$s_{\log}(m,n):=\log\left(s(m,n)\right)$ 
as
\begin{align}
    &\nabla\phi(m,n) =
    \left[\phi_\sfreq(m,n),\phi_\stime(m,n) \right] := \\
    & \left[ -\frac{\lambda }{a} (s_{\log}D_{\stime}^T)(m,n) ,
    \frac{1}{\lambda b} (D_{\sfreq}s_{\log})(m,n) + 2\pi m/M
\right]
\label{eq:numgrad}
\end{align}
where $D_{\stime}^T,D_{\sfreq}$ denote matrices performing the numerical differentiation
of $s_\text{log}$ 
along rows (in time) and columns (in frequency) respectively.
The matrices are assumed to be scaled
such that the sampling step of the differentiation scheme they represent is 
equal to 1.
The central (mid-point) finite difference scheme (see e.g. \cite{numericalanalysis10})
is the most suitable because it 
ensures the gradient components to be sampled at the same grid.

The steps of the numerical integration will be done in either horizontal or vertical
directions such that exclusively one of the components in
$\od{r}{\stimearg}$ from \eqref{eq:lineint} is zero.
Due to this property, the gradient can be pre-scaled
using lengths of the steps (hop sizes $a$ and $b$) such that 
\begin{align}
    &\nabla\phi^{\text{SC}}(m,n) :=  \left[ b \phi_\sfreq(m,n), a \phi_\stime(m,n) \right] =\\
    & \left[ -\frac{\lambda L}{a M} (s_{\log}D_{\stime}^T)(m,n) ,
    \frac{aM}{\lambda L} (D_{\sfreq}s_{\log})(m,n) + 2\pi am/M
\right].
\label{eq:numgradsc}
\end{align}
Note that the dependency on $L$ can be avoided
when \eqref{eq:gausssamples} is used to express $\lambda L$.
This is useful e.g. when the signal length is not known in advance.

The numerical gradient line integration is performed adaptively using the
simple trapezoidal rule.
The algorithm makes use of a heap data structure (from the heapsort algorithm
\cite{wi64}).
In case of the present algorithm it is used for holding pairs $(m,n)$ 
and it has the property of having $(m,n)$ of the maximum $|c(m,n)|$ always at the top.
It is further equipped with efficient operations for insertion and deletion.
The effect of the parameter $\mathit{tol}$ is twofold. 
First, a random phase (uniformly distributed random values from the range $\left[0,2\pi\right]$) is
assigned to coefficients small in magnitude for which the phase gradient is unreliable \cite{xxlbayjailsoend11}
and second, the integration is done only locally on ``islands'' with the
max coefficient within the island serving as the zero phase reference.
The randomization of the phase of the coefficients below the tolerance
is chosen over the zero phase because in practice it helps to avoid the impulsive
disturbances introduced by the small phase-aligned coefficients. 
The algorithm is summarized in Alg.~\ref{alg:heapint}.


After $\widehat{\phi}(m,n)$ has been estimated by Alg.~\ref{alg:heapint}, it is combined with the
target magnitude of the coefficients such that
\begin{equation}
    \widehat{c}(m,n)=s(m,n) \me^{\mi\widehat{\phi}(m,n)}
\end{equation}
and the signal is recovered by simply plugging these coefficients into \eqref{eq:idgt}.

\begin{algorithm}[th]

\ifCLASSOPTIONtwocolumn
\else
\footnotesize
\fi
    \LinesNumbered
    \caption{Phase gradient heap integration -- PGHI}
    \label{alg:heapint}
    \KwIn{DGT phase gradient
        $\nabla\phi^{\text{SC}}(m,n)=\left(\phi^{\text{SC}}_\sfreq(m,n),\phi^{\text{SC}}_\stime(m,n) \right)$ obtained
        from \eqref{eq:numgradsc},
        magnitude of DGT coefficients
    $\left|c(m,n)\right|$, relative tolerance $\mathit{tol}$.}
    \KwOut{Estimate of the DGT phase $\widehat{\phi}(m,n)$.}
    Set $\mathcal{I}=\left\{(m,n): \left|c(m,n)\right|>\mathit{tol} \cdot \max
    \left(\left|c(m,n)\right|\right) \right\}$\;
    Assign random values to $\widehat{\phi}(m,n)_{(m,n)\notin \mathcal{I} }$\;
    Construct a self-sorting \emph{heap} for $(m,n)$ pairs\;\label{alg:line}
    \While{$\mathcal{I}$ is not $\emptyset$ \label{alg:mark} }{
        \If{heap is empty}{
            Insert $(m,n)_{\text{max}} = \argmax
            \left(\left|c(m,n)_{(m,n)\in\mathcal{I}}\right|\right)$ into the \emph{heap}\;
            $\widehat{\phi}(m,n)_{\text{max}} \leftarrow 0$\;
            Remove $(m,n)_{\text{max}}$ from $\mathcal{I}$\;
        }
        \While{heap is not empty}{
            $(m,n) \leftarrow$ remove the top of the \emph{heap}\;
            \If{$(m+1,n) \in \mathcal{I}$}{
                $\widehat{\phi}(m+1,n) \leftarrow$ \\ \nonl\hfill $\widehat{\phi}(m,n) +
                \frac{1}{2}\left( \phi^{\text{SC}}_\sfreq(m,n) + \phi^{\text{SC}}_\sfreq(m+1,n)  \right)$\;
                Insert $(m+1,n)$ into the \emph{heap}\;
                Remove $(m+1,n)$ from  $\mathcal{I}$\;
            }
            \If{$(m-1,n) \in \mathcal{I}$}{
                $\widehat{\phi}(m-1,n) \leftarrow $ \\ \nonl\hfill $\widehat{\phi}(m,n) -
                \frac{1}{2}\left( \phi^{\text{SC}}_\sfreq(m,n) + \phi^{\text{SC}}_\sfreq(m-1,n)  \right)$\;
                Insert $(m-1,n)$ into the \emph{heap}\;
                Remove $(m-1,n)$ from  $\mathcal{I}$\;
            }
            \If{$(m,n+1) \in \mathcal{I}$}{
                $\widehat{\phi}(m,n+1) \leftarrow $ \\ \nonl\hfill $\widehat{\phi}(m,n) +
                \frac{1}{2}\left( \phi^{\text{SC}}_\stime(m,n) + \phi^{\text{SC}}_\stime(m,n+1)  \right)$\;
                Insert $(m,n+1)$ into the \emph{heap}\;
                Remove $(m,n+1)$ from  $\mathcal{I}$\;
            }
            \If{$(m,n-1) \in \mathcal{I}$}{
                $\widehat{\phi}(m,n-1) \leftarrow $ \\ \nonl\hfill $\widehat{\phi}(m,n) -
                \frac{1}{2}\left( \phi^{\text{SC}}_\stime(m,n) + \phi^{\text{SC}}_\stime(m,n-1)  \right)$\;
                Insert $(m,n-1)$ into the \emph{heap}\;
                Remove $(m,n-1)$ from  $\mathcal{I}$\;
            }
        }
    }
\end{algorithm}

\subsection{Practical Considerations}
\label{sec:practical}
In this section, we analyze the effect of the discretization
on the performance of the algorithm.

The obvious sources of error are the numerical differentiation and
integration schemes. However, the aliasing introduced by
subsampling in time and frequency domains is more serious.
In the discrete time setting, since the signal is considered to be band-limited and
periodic, the truly aliasing-free case occurs when $a=1, b=1$ ($M=L, N=L$)
regardless of the time or the frequency effective supports of the window.
DGT with such setting is however highly redundant and only signals up to several
thousands samples in length can be handled effectively.

In the subsampled case, the amount of aliasing and therefore the performance of
the algorithm depends on the effective support of the window.
Increasing $a$ introduces aliasing in frequency and increasing $b$
introduces aliasing in time.

\begin{figure}[thpb]
    \centering
    \subfloat[Spectrogram, $a=1$]{%
        \includegraphics[width=0.49\linewidth]{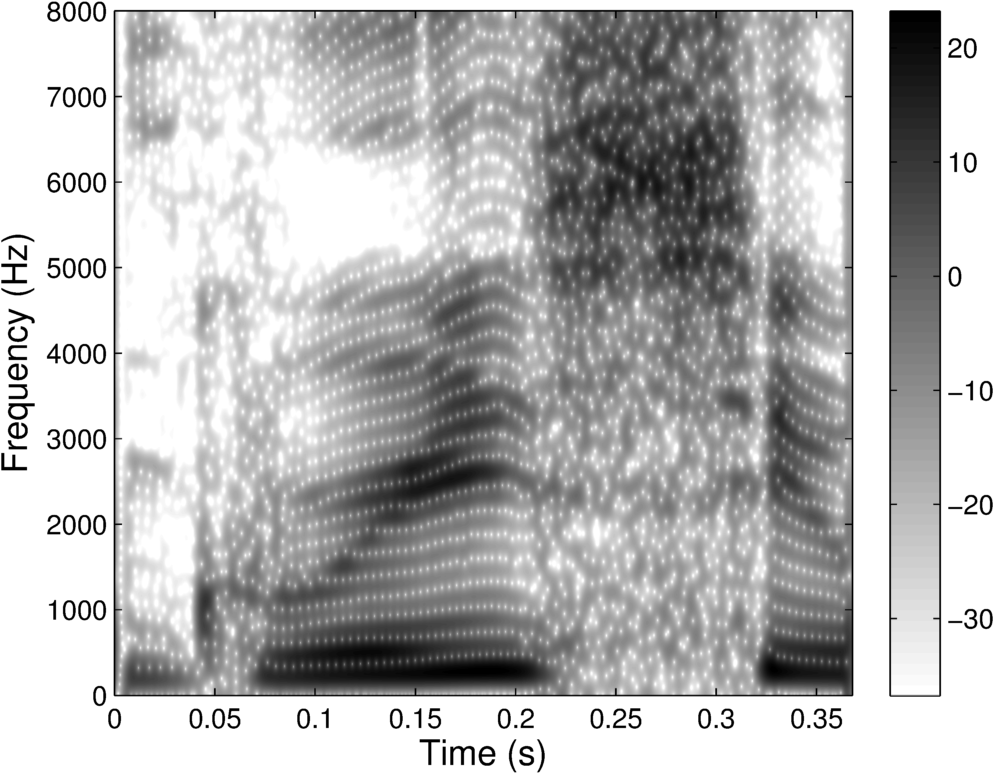}
        \label{subfig:greasy}
    }%
    \subfloat[][$a=\input{img/phasediffa_1.tex}$,
        $\mathcal{C}_\text{dB}=\input{img/phasediffC_1.tex}$
    ]{%
        \includegraphics[width=0.48\linewidth]{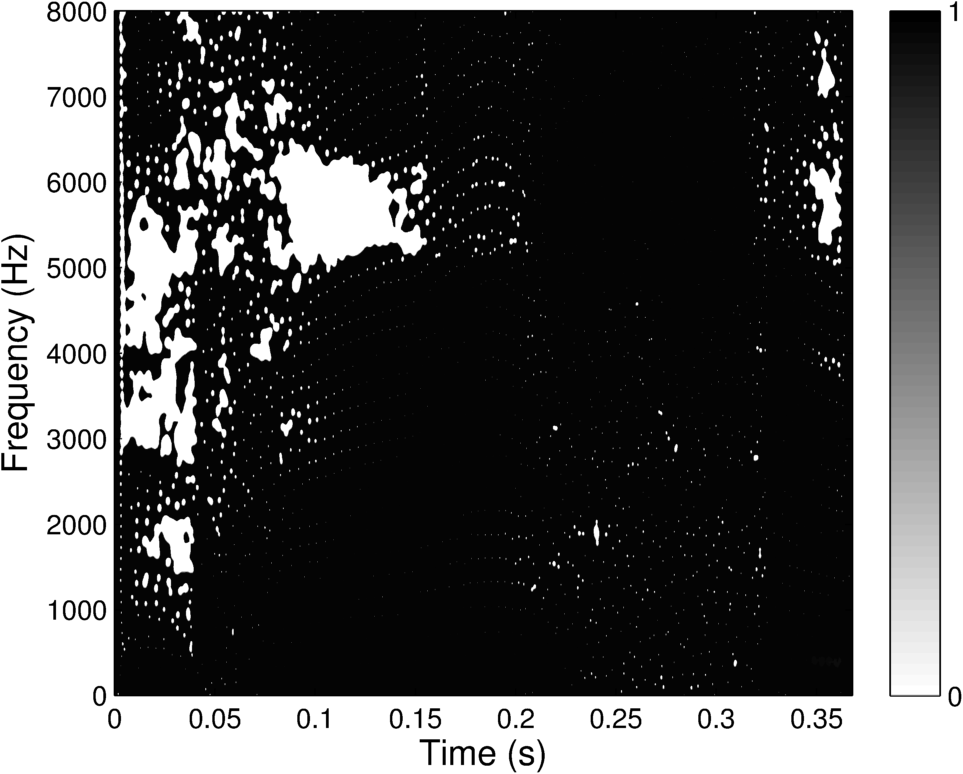}
        \label{subfig:a1}
    } \\ \vspace{-1em}
    \subfloat[][$a=\input{img/phasediffa_2.tex}$,
        $\mathcal{C}_\text{dB}=\input{img/phasediffC_2.tex}$
    ]{%
        \includegraphics[width=0.48\linewidth]{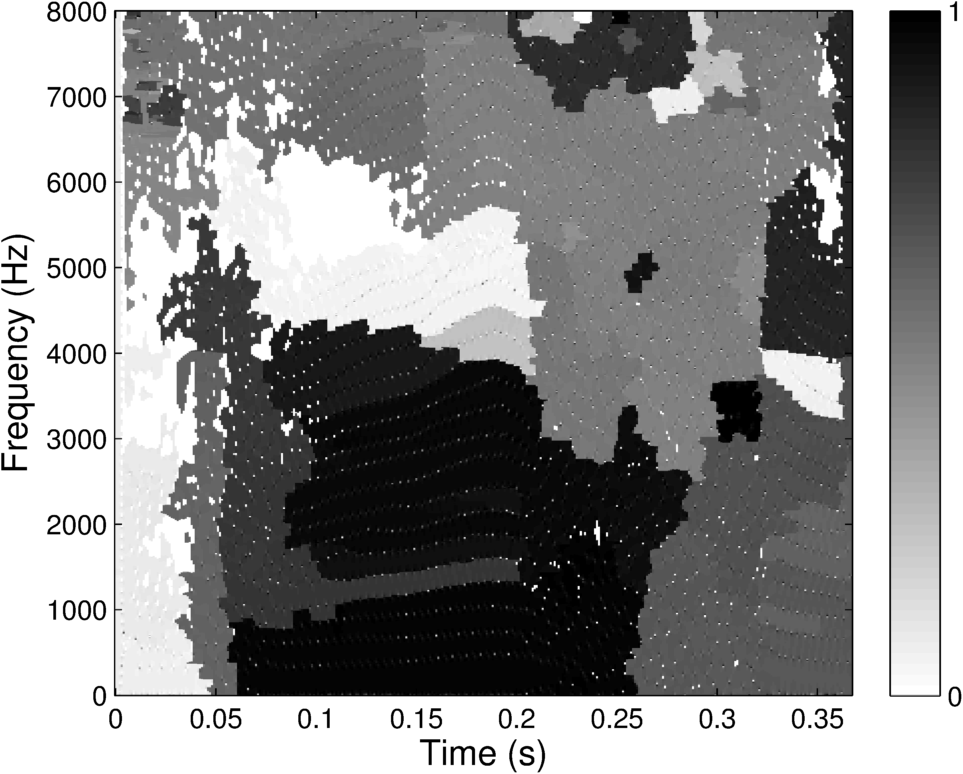} \
        \label{subfig:amore}
    }
    \subfloat[][$a=\input{img/phasediffa_3.tex}$,
        $\mathcal{C}_\text{dB}=\input{img/phasediffC_3.tex}$
    ]{%
        \includegraphics[width=0.48\linewidth]{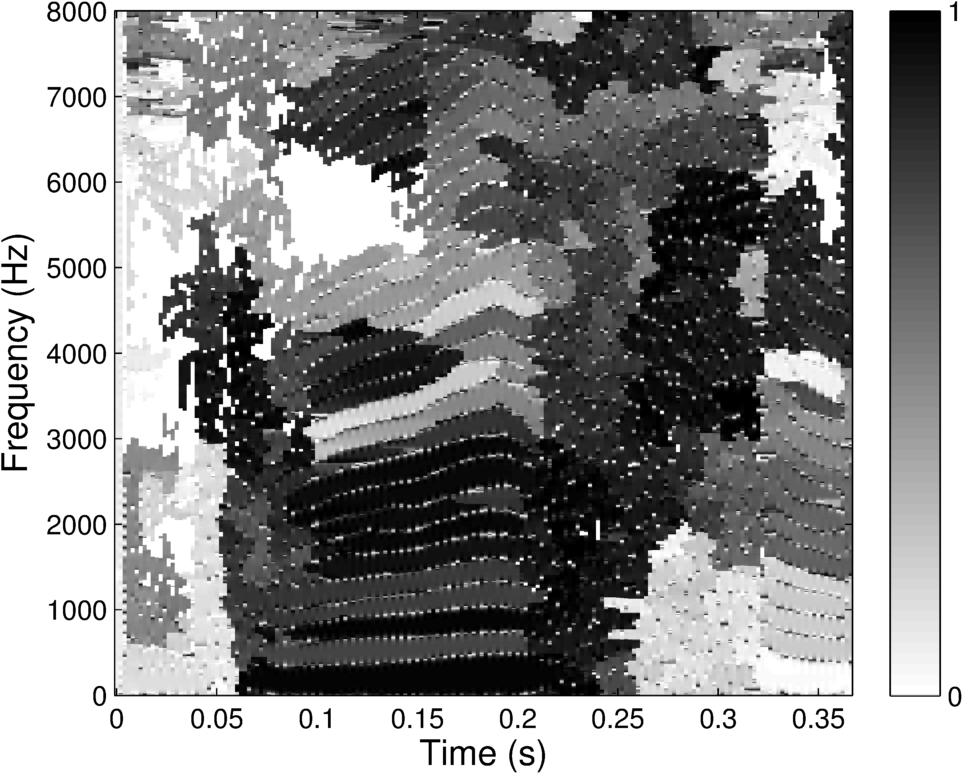}
        \label{subfig:aevenmore}
    }%

    \caption{%
         Spectrogram of a spoken word \emph{greasy} (a).  The absolute phase differences
     of the STFT of the original and reconstructed signal in $\text{rad}/\pi$ (modulo $1$)
 for varying  time hop size $a$ (b) (c) (d). The error $\mathcal{C}_\text{dB}$ 
 is introduced in Section~\ref{sec:exp}.}
    \label{fig:phasediff1}
\end{figure}

This property is illustrated by Figure~\ref{fig:phasediff1}, which shows that 
the algorithm performs very well in the
aliasing-free case ($a=1, b=1$) and the performance becomes worse when longer hop
sizes in time are introduced while keeping the effective width of the window 
constant. The length of the signal is 5888 samples and the
time-frequency ratio of the Gaussian window is $\lambda=1$.
The hop size in frequency is $b=1$ (i.e. $M=5888$).
Only the values for the 60 dB range of the highest coefficients are shown.

Even though the Gaussian window is in theory infinitely supported in both time and
frequency,
it decays exponentially and therefore aliasing might not
significantly degrade the performance of the algorithm
when choosing the hop sizes and the effective
support carefully. Obviously the finer the hop sizes the higher the computational cost.
The settings used in Section~\ref{sec:exp}, 
i.e. window overlap 87.5\% and overall redundancy 8 seem to be a good compromise.  

Since it is clear that the phase shift achieved by the algorithm is not constant,
the conjugate symmetry of the DGT of real signals cannot be easily recovered.
Therefore, when dealing with real
signals, we reconstruct the phase only for the positive frequency coefficients
and enforce the conjugate symmetry to the negative frequency coefficients.

\subsection{Exploiting Partially Known Phase}
\label{sec:partiallyknown}
In some scenarios, the true phase of some of the coefficients or
regions of coefficients is available. In order to exploit such information, the
proposed algorithm has to be adjusted slightly.
First, we introduce a mask to select the reliable coefficients and second, we
select the border coefficients i.e. coefficients with at least one neighbor
in the time-frequency plane with unknown phase.
Then we simply initialize the algorithm
with the border coefficients stored in the heap.

Formally, Algorithm \ref{alg:heapint} will be changed such that
steps summarized in Algorithm \ref{alg:heapintpartial} are inserted 
after line \ref{alg:line}.

\begin{algorithm}[th]
    \SetKwInput{KwIn}{Additional input}
    \LinesNumbered
    \caption{Initialization for partially known phase}
    \label{alg:heapintpartial}
    \KwIn{Set of indices of coefficients $\mathcal{M}$ with
    known phase $\phi(m,n)_{(m,n)\in\mathcal{M}}$. }
    $\widehat{\phi}(m,n) \leftarrow \phi(m,n)$ for $(m,n)\in\mathcal{M}$\;
    \For{$(m,n) \in \mathcal{M} \cap \mathcal{I}$ }{
        \If{ $(m+1,n)\notin\mathcal{M}$ or $(m-1,n)\notin\mathcal{M}$ or
            $(m,n+1)\notin\mathcal{M}$ or $(m,n-1)\notin\mathcal{M}$
        }{
            Add $(m,n)$ to the \emph{heap}\;
        }
    }

\end{algorithm}

Note that the phase of the border coefficients can be used directly
(i.e. no unwrapping is necessary). Depending on the situation, the phase might be
propagated from more than one border coefficient, however the phases coming
from distinct sources are never combined.

\subsection{Connections to Phase Vocoder}
\label{sec:vocoder}
In this section we discuss some connections between the proposed algorithm and
the phase vocoder \cite{lado99} and consequently algorithms presented in
\cite{be15} and \cite{mabada15}.

The phase vocoder allows to change the signal duration by employing non-equal analysis and synthesis 
time hop sizes. A pitch change
can be achieved by playing the signal at a sampling rate adjusted by
the ratio of the analysis and synthesis hop sizes. In the synthesis, the phase must be
kept \emph{consistent} in order not to introduce artifacts.
In the phase reconstruction
task, the original phase is not available, but
the basic phase behavior can be yet exploited.
For example, it is known that for a sinusoidal component with a constant frequency the phase
grows linearly in time for all frequency channels the component influences in the spectrogram.
For these coefficients, the instantaneous frequency
(STFT phase derivative with respect to time \eqref{eq:phit}) is
constant and the local group delay (STFT phase derivative with respect to
frequency \eqref{eq:phiomega}) is zero.

In the aforementioned papers \cite{be15} and \cite{mabada15}, the instantaneous frequency is
estimated in each spectrogram column (time frame) from the magnitude by peak picking
and interpolation. The instantaneous frequency determines phase increments
for each frequency channel $m$ such that
\begin{equation}\label{eq:phasetupdate}
    \phi(m,n) = \phi(m,n-1) + 2\pi a m_0 /M,
\end{equation}
where $m_0$ is the estimated, possibly non-integer instantaneous frequency 
belonging to the interval $\left[0,\lfloor M/2\rfloor\right]$.
This is exactly what the proposed algorithm does in case of constant sinusoidal
components, except the instantaneous frequency
is determined from the DGT log-magnitude. 
Integration in Alg.~\ref{alg:heapint} performs nothing else than a cumulative
sum of the instantaneous frequency in the time direction.

The algorithm from \cite{mabada15} goes further and employs an impulse model.
The situation is reciprocal to sinusoidal components such that the phase
changes linearly in frequency for all coefficients belonging
to an impulse component but the rate is only constant for
fixed $n$ and it is inversely proportional to the local group delay $n_0-n$
 such as
\begin{equation}\label{eq:phasefupdate}
    \phi(m,n) = \phi(m-1,n) + 2\pi a ( n - n_0) /M,
\end{equation}
where $an_0$ is the time instant of the impulse occurrence. 
Again, this is what the proposed algorithm does for coefficients corresponding to
impulses.

The advantage of the proposed algorithm over the other two is that the phase
gradient is computed from the DGT log-magnitude such that it is available
at every time-frequency position
without even analysing the spectrogram
content. This allows an arbitrary integration path which combines both
the instantaneous frequency and the local group delay according to the
magnitude ridge orientation.
In the other approaches, the phase time
derivative can be only estimated in a vicinity of
sinusoidal components and, vice versa, the
frequency derivative only in a vicinity of impulse-like
events.
Obviously, such approaches will not cope well with
deviations from the model assumptions although
careful implementation can handle multiple sinusoidal components with slowly
varying instantaneous frequencies and impulses with frequency-varying onsets.
The difficulty of algorithm \cite{mabada15} lies in detecting
the onsets in the spectrogram and separating the coefficients belonging
to the impulse-like component from the coefficients
belonging to sinusoidal components.

\begin{figure}[t]
    \centering
    \subfloat[Spectrogram]{%
        \includegraphics[width=0.5\linewidth]{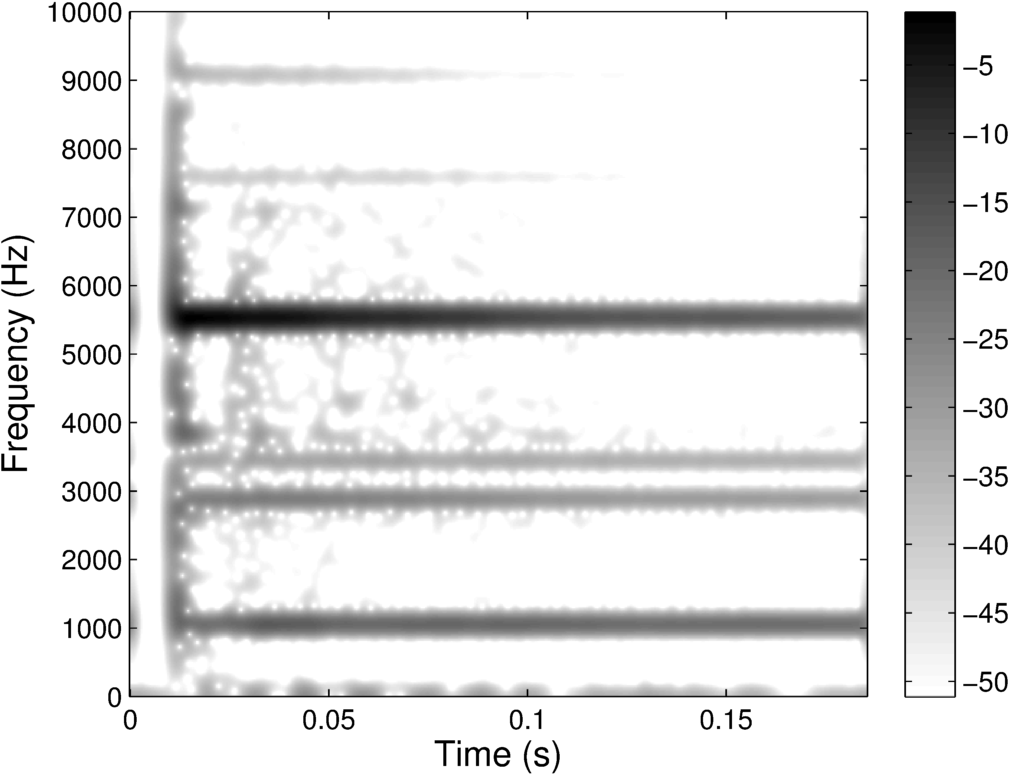}
        \label{subfig:gspi}
    }
    \subfloat[][Alg. \cite{be15}%
    ]{%
        \includegraphics[width=0.48\linewidth]{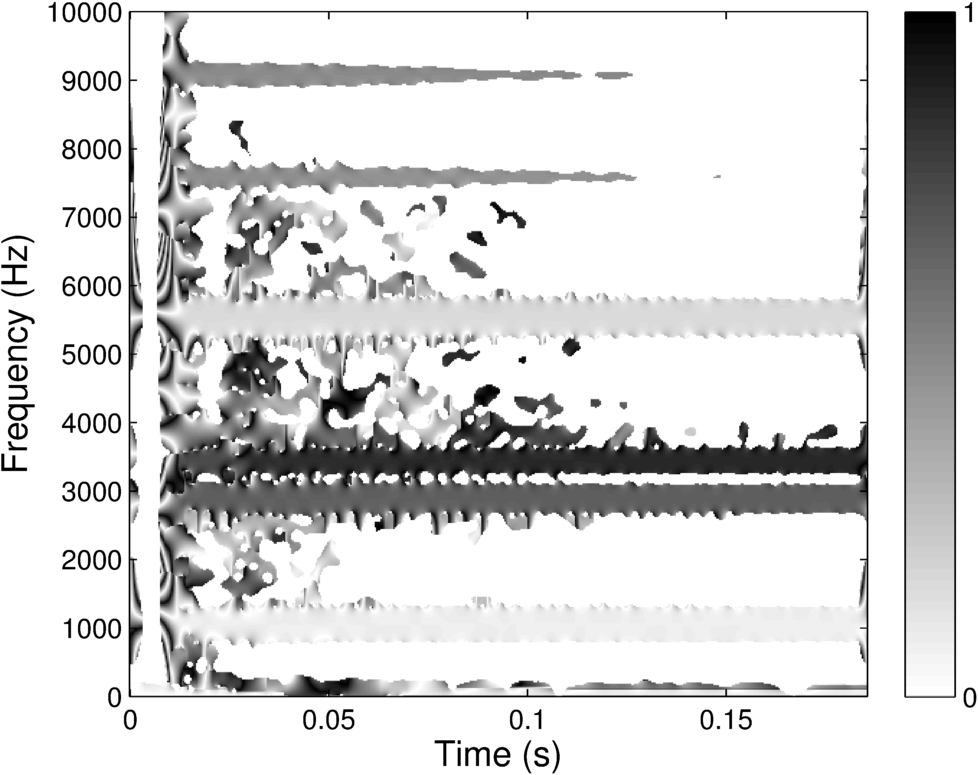}
        \label{subfig:gspispsi}
    } \\ \vspace{-1em}
    \subfloat[][Alg. \cite{mabada15}%
    ]{%
        \includegraphics[width=0.48\linewidth]{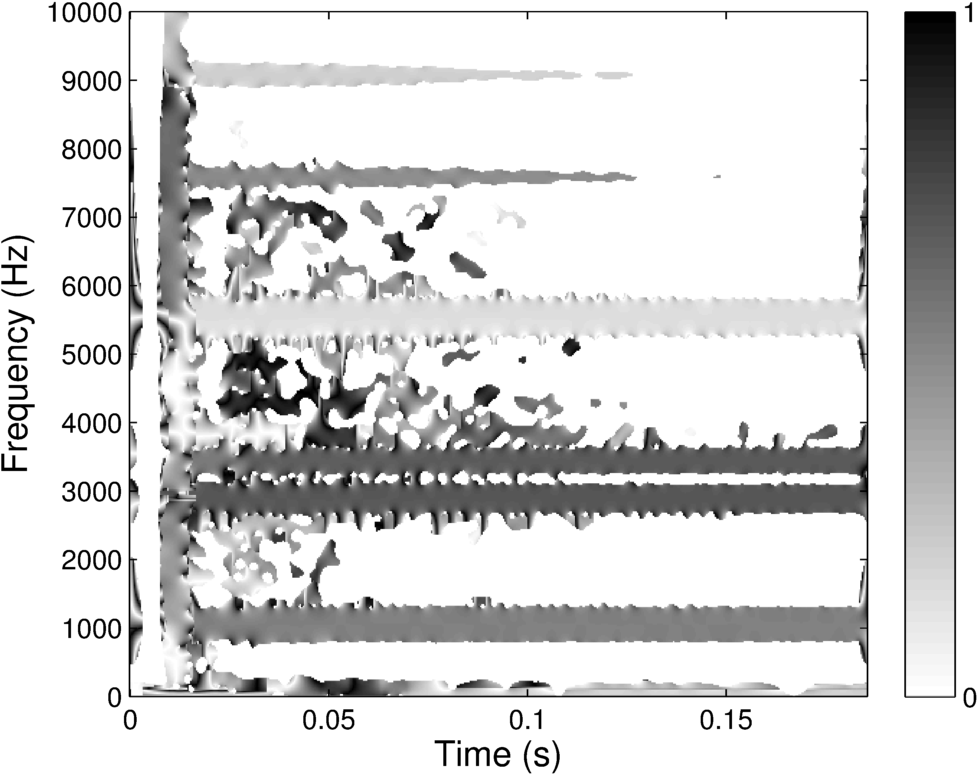} \
        \label{subfig:gspiunwrap}
    }
    \subfloat[][Proposed
    ]{%
        \includegraphics[width=0.48\linewidth]{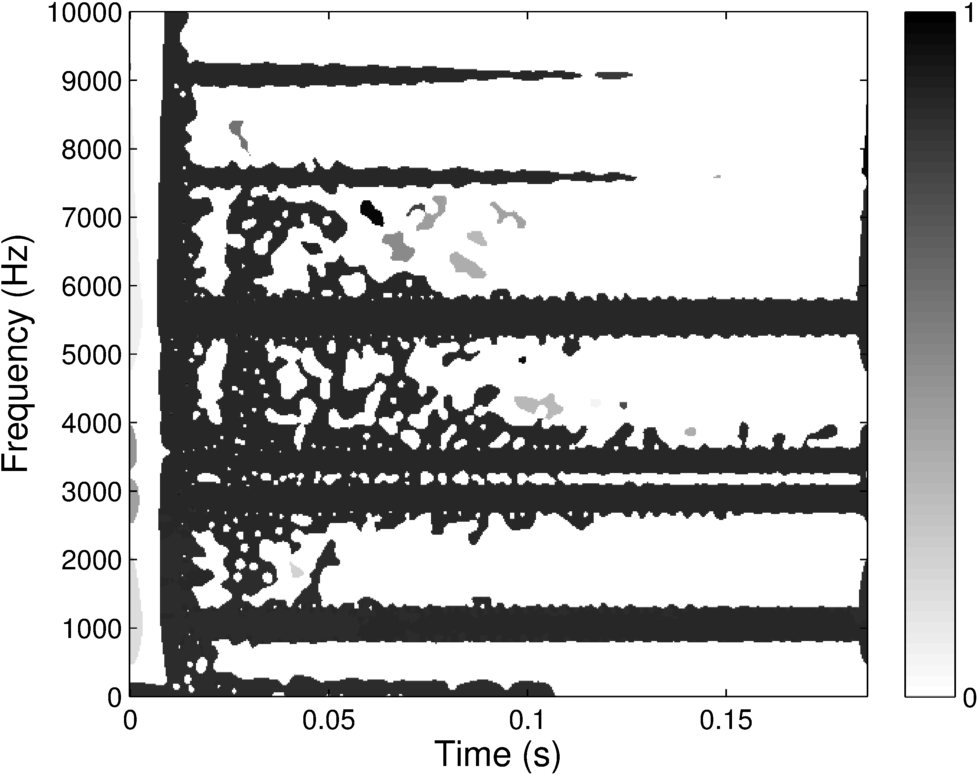}
        \label{subfig:gspiheapint}
    }%

    \caption{%
        Spectrogram of an excerpt form the \emph{glockenspiel} signal and
        the absolute phase differences in $\text{rad}/\pi$ (modulo $1$) for
        three different algorithms.
}
    \label{fig:phasediff2}
\end{figure}

Figure \ref{fig:phasediff2} shows phase deviations achieved by algorithms
from \cite{be15}, \cite{mabada15} and by the proposed algorithm.
The phase difference at the transient coefficients is somewhat smoother
for Alg. \cite{mabada15} when compared to \cite{be15} because of the
involved impulse model.
The proposed algorithm produces almost constant phase difference due to the
adaptive integration direction.
The setup used in the example is the following: 
the length of the signal is $L=8192$ samples, time hop size $a=16$,
number of channels $M=2048$, time-frequency ratio of the Gaussian
window is $\lambda=aM/L$.
Only the values for the 50 dB range of the highest
coefficients are shown.

\section{Experiments}
\label{sec:exp}

In the experiments, we use the following equation to measure
the error
\begin{equation}
    E(x,y) = \frac{\left\lVert x-y\right\rVert_{\text{2}}}{\left\lVert
    x\right\rVert_{\text{2}}},\ 
    E_{\text{dB}}(x,y)=20\log_{10}E(x,y),
\end{equation}
where $x,y$ are either vectors or matrices \xxl{ and $\left\lVert . \right\rVert_{\text{2}}$ denotes the standard energy norm.} 
In \cite{stda11} the \emph{spectral convergence} is defined as
\begin{equation}
    \mathcal{C}
    =  E\left(s,\left|P\widehat{c}\right|\right),\
    \mathcal{C}_{\text{dB}}(x,y)= 20\log_{10}\mathcal{C},
    \label{eq:consistency}
\end{equation}
where $P = F^*_gF_{\widetilde{g}}$.
In \cite{leroux10} the authors proposed a slightly
different measure
$E(\widehat{c},P\widehat{c})^2$
called \emph{normalised inconsistency measure}
defining normalised energy lost by the reconstruction/projection.
Such measures clearly do not accurately
reflect the actual signal reconstruction error $E(f,\widehat{f})$,
but they are independent of the phase shift.
Some other papers evaluate the algorithms using the signal to noise ratio,
which they define as $\mathit{SNR}(x,y)=1/E(x,y)$ and $\mathit{SNR}_{\text{dB}}(x,y)=-E_{\text{dB}}(x,y)$ respectively. 

Unfortunately, as the
phase difference plots in Fig.\,\ref{fig:phasediff1} and Fig.\,\ref{fig:phasediff2} show,
the phase difference is usually far from
being constant when subsampling is involved 
(this holds for any algorithm, even the iterative ones).
Therefore, the time-frames (i.e. individual short-time spectra) and even each
frequency bin within the frame
might have a different phase shift, causing the error $E(f,\widehat{f})$
to be very high, even when the other error measures are low and the actual
perceived quality is good.

The testing was performed on the speech corpus database
MOCHA-TIMIT \cite{mochatimit} consisting of recordings of 1 male and 1 female
English speakers each of which performing 460 sentences. The total duration of the
recordings is 61 minutes and 5 seconds.
The sampling rate of all recordings is 16\,kHz.
The Gabor system parameters used with this database 
(Table\,\ref{tab:speech} and Fig.\,\ref{fig:compareallmocha})
were:
number of channels $M=1024$, hop size $a=128$, time-frequency ratio
of the Gaussian window $\lambda=aM/L$, time support of the truncated
Gaussian window and the other compactly supported windows was $M$ samples.

Next, we used the EBU SQAM database of 70 test sound samples \cite{sqam}
recorded at 44{.}1 kHz. Only the first 10 seconds of the first channel was
used from the stereophonic recordings to reduce the execution time to a
reasonable value. 
The Gabor system parameters used with this database
(Table\,\ref{tab:music} and Fig.\,\ref{fig:compareallsqam}) were the following:
number of channels $M=2048$, hop size $a=256$, time-frequency ratio
of the Gaussian window $\lambda=aM/L$, time support of the truncated
Gaussian window and of the other compactly supported windows was $M$ samples.


\subsection{Performance Of The Proposed Algorithm}
\label{sec:perf}
In this section, we evaluate the performance of the proposed algorithm
and compare it to results obtained by the Single Pass Spectrogram Inversion
algorithm \cite{be15} (SPSI).
Unfortunately, we were not able to get good
results with the algorithm from \cite{mabada15} consistently due to the
imperfect onset detection and due to the limitation of the impulse model \xxl{and so did not include it here.}

The implementation of SPSI has been taken from
\url{http://anclab.org/software/phaserecon/}
and it was modified to fit our framework. The most prominent change has been the
removal of the alternating $\pi$ and $0$ phase modulation in the frequency direction
which is not present when computing the transform according to \eqref{eq:dgt}.


The results for the proposed algorithm were computed via a two step procedure.
In the first step Alg.~\ref{alg:heapint} with $\mathit{tol}=10^{-1}$ was used,
and in the second step, the algorithm was run again with $\mathit{tol}=10^{-10}$
including steps from Alg.~\ref{alg:heapintpartial} while using 
the result from the first step as known phase.

This approach avoids error spreading during the numerical integration and improves
the result considerably when compared to a single run with either of the thresholds.

Tables\,\ref{tab:speech} and \ref{tab:music} show the average $\mathcal{C}$ 
(converted to a value in dB)
over whole databases for the SPSI and the proposed algorithm.
The proposed algorithm very clearly outperforms the SPSI algorithm by a large margin.
The performance of the proposed algorithm further depends on the choice of the window. 
While the Gaussian window truncation introduces only a negligible 
performance degradation, the choice of Hann or Hamming windows increase
the error by about 2 dB.
For a detailed comparison, please find the scores and sound examples 
for the individual files from the EBU SQAM database using the Gaussian window
at the accompanying web page \url{http://ltfat.github.io/notes/040}.

\begin{table}[!tpb]
    \centering
    \caption{Average $\mathcal{C}$ in dB for the MOCHA-TIMIT database}
    \label{tab:speech}
    \begin{tabular}{c|c|c|c|c}
        & Gauss & Trunc. Gauss & Hann & Hamming \\ \hline \hline
        SPSI \cite{be15}&%
        $\input{img/spsigaussmocha}$&
        $\input{img/spsitruncgaussmocha}$&
        $\input{img/spsihannmocha}$&
        $\input{img/spsihammingmocha}$
        \\
        PGHI (proposed)&%
        $\mathbf{\input{img/heapintgaussmocha}}$&
        $\mathbf{\input{img/heapinttruncgaussmocha}}$&
        $\mathbf{\input{img/heapinthannmocha}}$&
        $\mathbf{\input{img/heapinthammingmocha}}$
        \\ \hline
    \end{tabular}
\end{table}

\begin{table}[!tpb]
    \centering
    \caption{Average $\mathcal{C}$ in dB for the EBU SQAM database}
    \label{tab:music}
    \begin{tabular}{c|c|c|c|c}
        & Gauss & Trunc. Gauss & Hann & Hamming \\ \hline \hline
        SPSI \cite{be15}&%
        $\input{img/spsigausssqam}$&%
        $\input{img/spsitruncgausssqam}$&%
        $\input{img/spsihannsqam}$&%
        $\input{img/spsihammingsqam}$%
        \\
        PGHI (proposed)&%
        $\mathbf{\input{img/heapintgausssqam}}$&%
        $\mathbf{\input{img/heapinttruncgausssqam}}$&%
        $\mathbf{\input{img/heapinthannsqam}}$&%
        $\mathbf{\input{img/heapinthammingsqam}}$%
        \\ \hline
    \end{tabular}
\end{table}

We can only provide a rough timing for the algorithms as the actual execution time
is highly signal dependent and our implementations might be suboptimal.
In the setup used in the tests, the runtime of the proposed algorithm is about
6--8 times longer than of the implementation of the SPSI algorithm.
In particular, the current implementation of the proposed algorithm is very
slow for noise signals.


\subsection{Comparison With The State-of-the-art}
We further compare the present algorithm with the following iterative algorithms:
\begin{itemize}
    \item The Griffin-Lim algorithm \cite{griflim84} (GLA) as the baseline.
    \item A combination of Le Roux's modifications of GLA \cite{leroux10}
        and the fast version of GLA \cite{pebaso13} with constant $\alpha=0.99$ (FleGLA).
        More precisely from \cite{leroux10} we use the modification called
        \emph{on-the-fly truncated modified update} which was
        reported to perform the best. The on-the-fly phase updates are performed
        in the natural order of frames starting with the zero frequency bin within each
        frame.
        The projection kernel was always truncated to size  $2M/a-1$ in both directions.

        This combination outperformes both algorithms \cite{leroux10} and \cite{pebaso13}
        when used individually.
    \item The gradient descend-like algorithm from \cite{desomada15} (lBFGS)
        with the refined objective function ($p=2/3$).
        Unfortunately, the lBFGS implementation we use (downloaded from \cite{minfunc}) fails
        in some cases.
    \item The real-time iterative spectrogram inversion algorithm 
        with look-ahead (RTISI-LA).
        The algorithm was published in \cite{zhbewy07}, but we implemented
        a refined version using the truncated projection kernel from \cite{leroux10}
        as proposed in \cite{leroux10b}. 
        The number of the look-ahead frames was always $M/a-1$ and an asymmetric
        analysis window was used for the latest look-ahead frame.
        Due to the nature of the algorithm, its performance can only be 
        evaluated at $M/a$ multiples of per-frame iterations.
\end{itemize}

Since all the iterative algorithms optimize a non-convex objective
function, the result depends strongly on the initial phase estimate.
In addition to the zero phase initialization, 
we also evaluate performance of the algorithms
initialized with the phase computed using the proposed algorithm.
We will denote such initialization as \emph{warm-start} (ws).
Unfortunately, our tests showed that the RTISI-LA algorithm
does not benefit from the warm-starting as it performs
its own initial phase guess from the partially reconstructed signal.


Figures~\ref{fig:compareallmocha} and \ref{fig:compareallsqam}
show average $\mathcal{C}$ in dB over the
MOCHA-TIMIT and EBU SQAM databases respectively depending on the number of
iterations without (solid lines) or with (dashed lines) the warm start.
The horizontal dashed line is the average $\mathcal{C}$ in dB achieved by the
proposed algorithm (values from Tables\,\ref{tab:speech} and \ref{tab:music}). 
In addition, the scores and sound examples for individual files from the EBU SQAM database using
the Gaussian window can be found at the accompanying web page.

Graphs for the truncated Gaussian window are not shown as they exhibit 
no visual difference from the graphs for the full-length Gaussian window.
Further, the lBFGS algorithm has been excluded from the comparison using the EBU-SQAM 
database (Fig.\,\ref{fig:compareallsqam});
it failed to finish for a considerable number of the excerpts.

The graphs show that the proposed algorithm provides a suitable initial phase 
for the iterative algorithms i.e. the best overall results are obtained 
when combined with the FleGLA and lBFGS algorithms.
When considering the iterative algorithms without warm-starting, the crossing
points indicating the number of iterations necessary to achieve the performance 
of the proposed algorithm can be clearly identified (if present).
For the MOCHA TIMIT database (Fig.~\ref{fig:compareallmocha}), the crossing 
point of the best algorithms is at about 20 iterations and the behavior
is consistent for all the windows.
The results for the EBU SQAM database (Fig.~\ref{fig:compareallsqam})
are more erratic. 
First of all, the GLA algorithm never reaches the performance of
the proposed algorithm even in 200 iterations. The crossing point of the
FleGLA algorithm varies from 45 to 170 iterations depending on the window used.
On the other hand, the RTISI-LA algorithm gets close to the line only for 8 per-frame
iterations using the Gaussian window (Fig.~\ref{fig:compareallsqama}) and
it performs better than the proposed algorithm in all the other cases.
The RTISI-LA algorithm however ``fails'' for sound excerpts
like castanets (crossing point at 80--120 iterations), drums, cymbals and glockenspiel 
(crossing points 20--40 iterations).

In the tests, the execution time of the proposed algorithm was comparable
to the execution time
of 2--4 iterations of the GLA algorithm with the Gaussian window and to
the execution time of 4--10 iterations for the compactly supported windows.

\ifCLASSOPTIONtwocolumn
\begin{figure}[t]
    \centering
    \subfloat[][Gaussian window]{\includegraphics[width=0.9\linewidth]{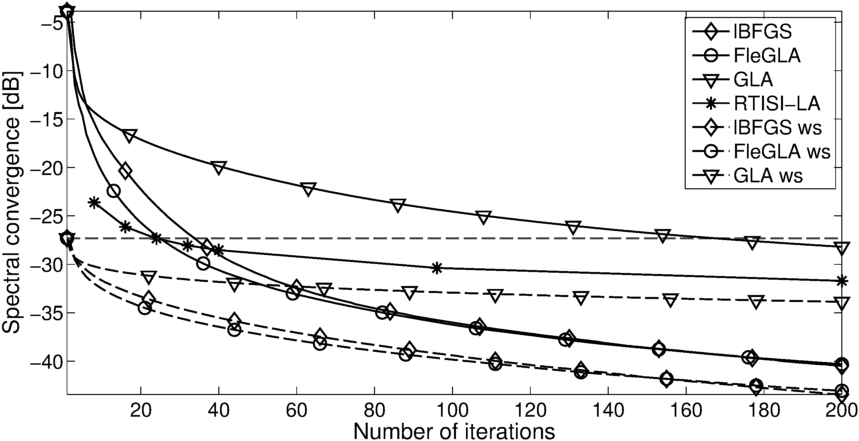}
    }\\
    \subfloat[][Hann window]{\includegraphics[width=0.9\linewidth]{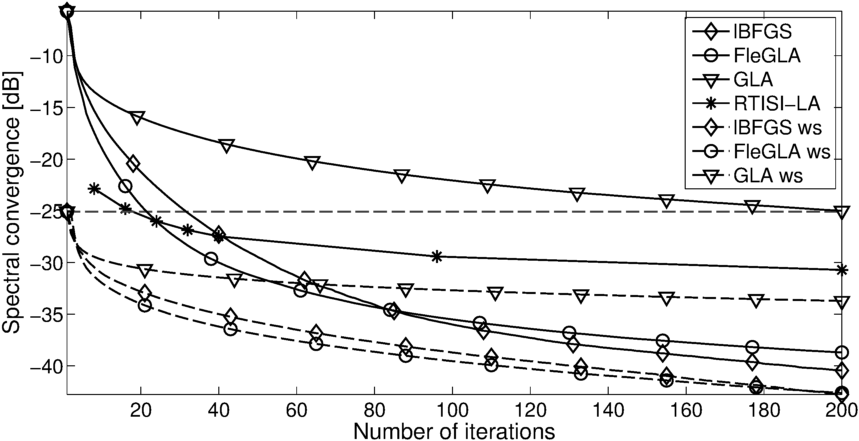}
    }\\
    \subfloat[][Hamming
    window]{\includegraphics[width=0.9\linewidth]{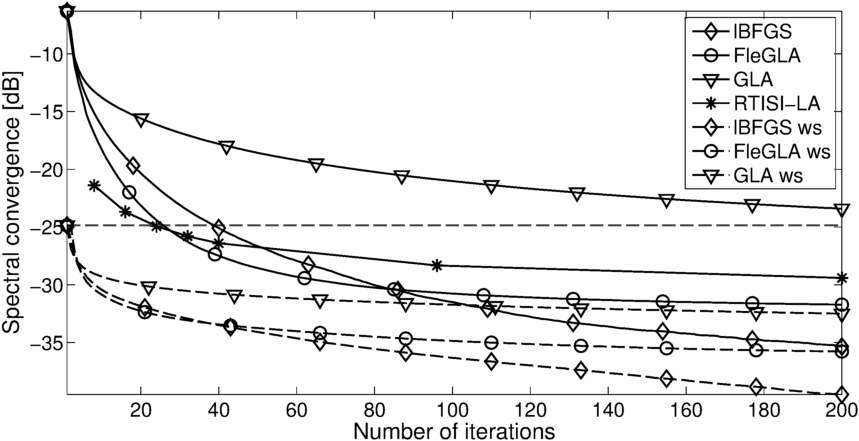}
    }
    \caption{Comparison with the iterative algorithms, MOCHA-TIMIT database. \xxl{(PGHI gives the horizontal dashed line.)}}
    \label{fig:compareallmocha}
\end{figure}

\begin{figure}[t]
    \centering
    \subfloat[][Gaussian
    window]{\includegraphics[width=0.9\linewidth]{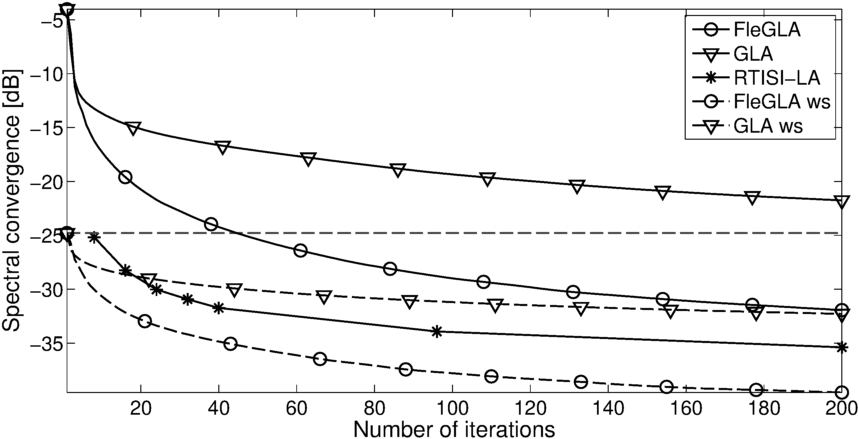}
    \label{fig:compareallsqama}}\\
    \subfloat[][Hann window]{\includegraphics[width=0.9\linewidth]{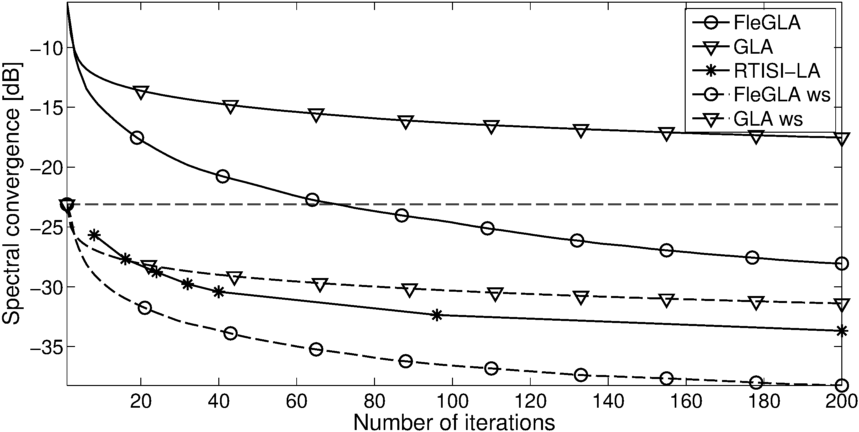} }\\
    \subfloat[][Hamming window]{\includegraphics[width=0.9\linewidth]{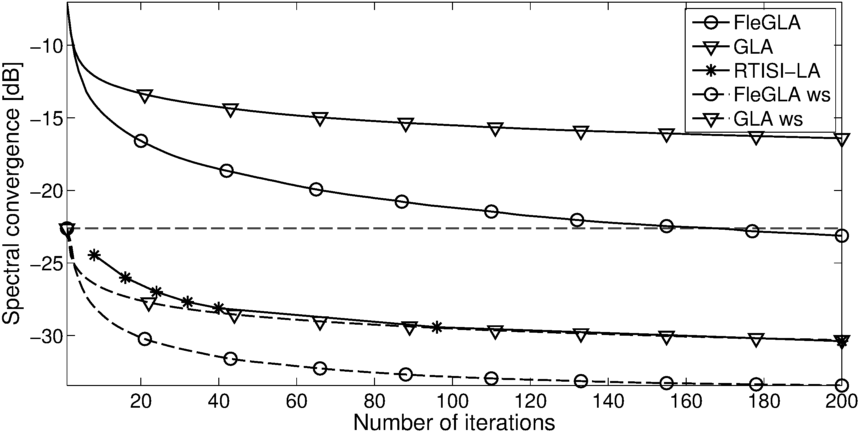} }
    \caption{Comparison with the iterative algorithms, EBU SQAM database. \xxl{(PGHI gives the horizontal dashed line.)}}
    \label{fig:compareallsqam}
\end{figure}
\else%
\begin{figure}[t]
    \centering
    \subfloat[][Gaussian window]{\includegraphics[width=0.8\linewidth]{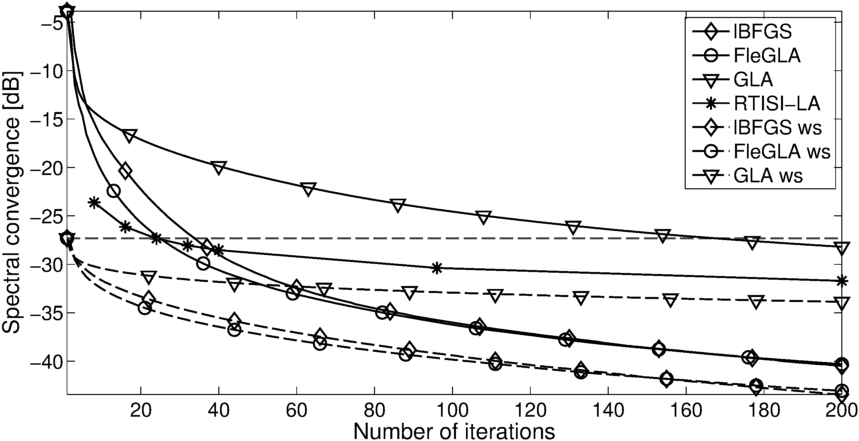}
    }\\
    \subfloat[][Hann window]{\includegraphics[width=0.8\linewidth]{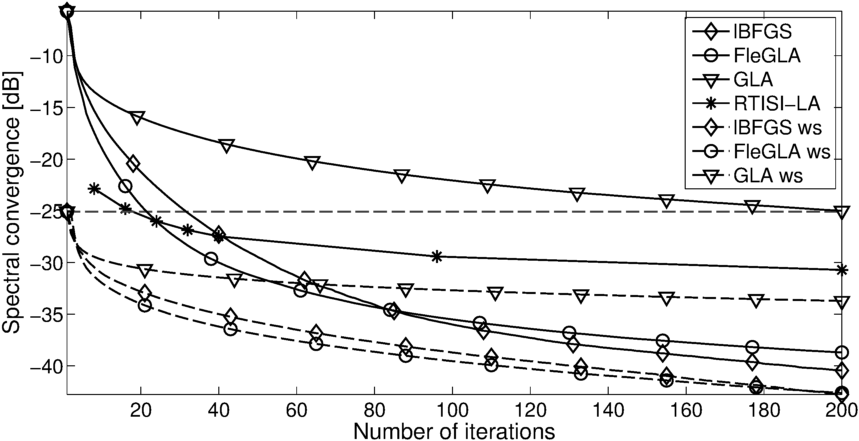}
    }\\
    \subfloat[][Hamming
    window]{\includegraphics[width=0.8\linewidth]{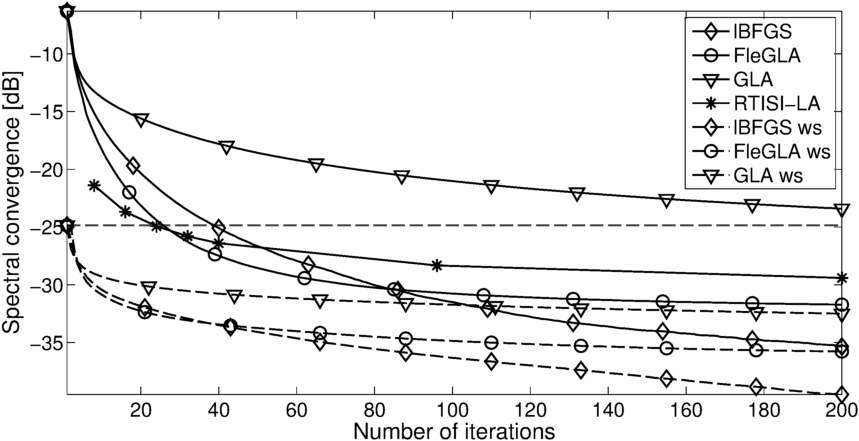}
    }
    \caption{Comparison with the iterative algorithms, MOCHA-TIMIT database. \xxl{(PGHI gives the horizontal dashed line.)}}
    \label{fig:compareallmocha}
\end{figure}

\begin{figure}[t]
    \centering
    \subfloat[][Gaussian
    window]{\includegraphics[width=0.8\linewidth]{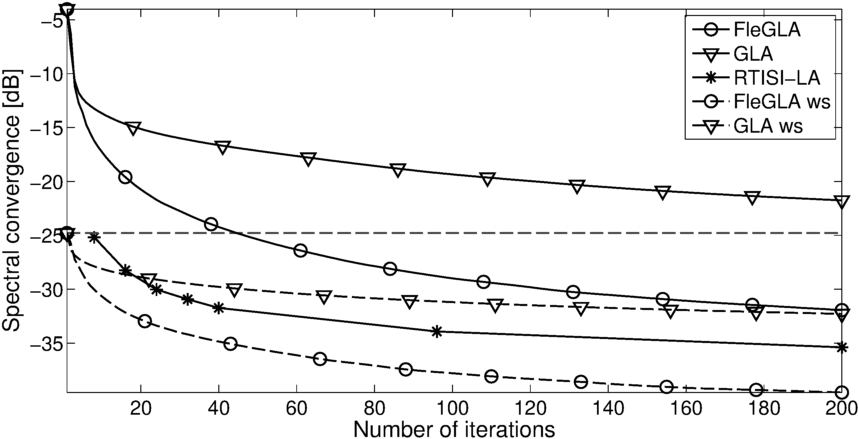}
    \label{fig:compareallsqama}}\\
    \subfloat[][Hann window]{\includegraphics[width=0.8\linewidth]{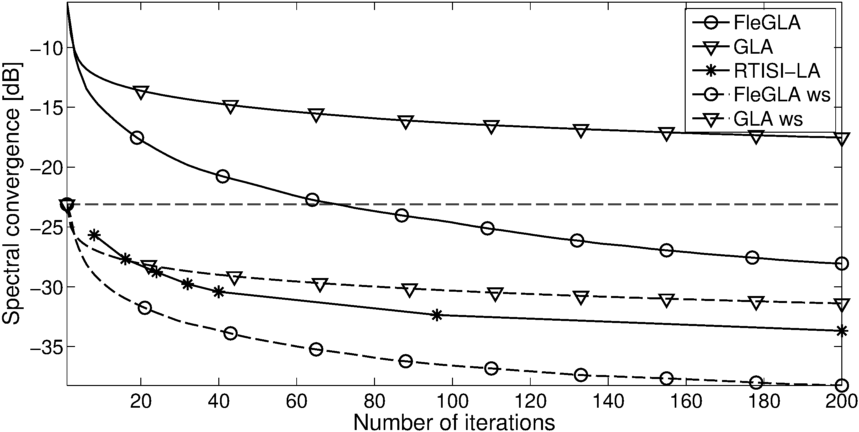} }\\
    \subfloat[][Hamming window]{\includegraphics[width=0.8\linewidth]{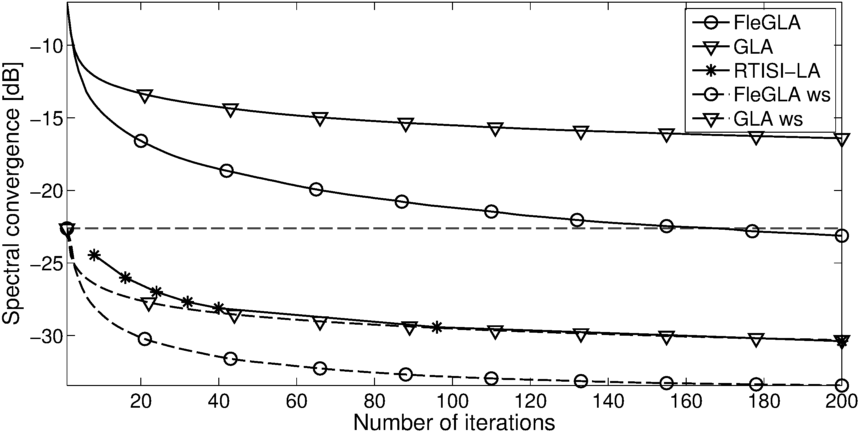} }
    \caption{Comparison with the iterative algorithms, EBU SQAM database. \xxl{(PGHI gives the horizontal dashed line.)}}
    \label{fig:compareallsqam}
\end{figure}

\fi

\subsection{Modified Spectrograms}
\label{sec:mstft}
The main application area of the phase reconstruction algorithms is
the reconstruction from the modified spectrograms. 
\xxl{The spectrograms are modified in the coefficent domain. This could be done by
multiplication which leads to so-called Gabor filters \cite{839987,xxllabmask1} or by moving/copying of contents.
In general, such a modified spectrogram is no longer a valid
spectrogram, i.e. there is no signal having such spectrogram. Therefore the task is to construct rather than
reconstruct a suitable phase.
Unfortunately, it is neither clear for which spectrogram modifications the equations
\eqref{eq:phiomega} and \eqref{eq:phit} still hold nor how it does affect the
performance if they do not. 
Moreover, an objective comparison of the algorithms becomes difficult as 
the error measures chosen above become irrelevant.}

Nevertheless, in order to get the idea of the performance of the proposed algorithm acting
on modified spectrograms, we implemented phase vocoder-like pitch shifting (up and down by 6
semitones) via changing the hop size (\cite{lado99,zo02}) using all the algorithms
to rebuild the phase. The synthesis hop size $a=256$ was fixed and the analysis hop
size was changed accordingly to achieve the desired effect. 
Sound examples for the EBU SQAM database along with Matlab/GNU
Octave script generating them can be found at the accompanying web page.
According to our informal listening tests, there is a little perceivable difference 
between the algorithms with the exception of SPSI. 
As expected, the SPSI algorithm introduces disturbing ``echo-like''
effects to sounds that do not conform with the model assumption.

\section{Conclusion}
A novel, non-iterative algorithm for the reconstruction of the phase from the STFT
magnitude has been proposed. The algorithm is computationally efficient and
its performance is competitive with the state-of-the-art
algorithms. It can also provide a suitable initial phase for the iterative algorithms.

As a future work it would be interesting to investigate whether (simple) equations similar
to \eqref{eq:phiomega} and \eqref{eq:phit} could be
found for non-Gaussian windows. 
Moreover, the effect of the aliasing and 
spectrogram modifications on the phase-magnitude relationship 
should be systematically explored.
\xxl{For that we will extend Proposition \ref{prop:gencauchy} to a more general setting. Ideally, we hope that a similar result could be possible for $\alpha$-modulation frames \cite{dahtes08,BaySpe2014} and warped time-frequency frames \cite{holwie15,holwiexxl15}.}

From the practical point of view, a drawback of the proposed algorithm is the
inability to run in real-time setting i.e. to process streams of audio data in a frame by frame
manner. Clearly, the way how the phase is spread among the coefficients would have to be
adjusted. This was done in \cite{ltfatnote043} where we present a version of the algorithm
introducing one or even zero frame delay.

\section*{Acknowledgements}
The authors thank Pavel Rajmic for his valuable comments.

%

%% file: img/phasediffa_1.tex
1

%% file: img/phasediffC_1.tex
-57.02

%% file: img/phasediffa_2.tex
16

%% file: img/phasediffC_2.tex
-28.17

%% file: img/phasediffa_3.tex
32

%% file: img/phasediffC_3.tex
-24.06

%% file: img/spsigaussmocha.tex
-16.75

%% file: img/spsitruncgaussmocha.tex
-16.75

%% file: img/spsihannmocha.tex
-14.53

%% file: img/spsihammingmocha.tex
-14.02

%% file: img/heapintgaussmocha.tex
-27.36

%% file: img/heapinttruncgaussmocha.tex
-27.37

%% file: img/heapinthannmocha.tex
-25.09

%% file: img/heapinthammingmocha.tex
-24.87

%% file: img/spsigausssqam.tex
-18.01

%% file: img/spsitruncgausssqam.tex
-18.19

%% file: img/spsihannsqam.tex
-17.09

%% file: img/spsihammingsqam.tex
-16.79

%% file: img/heapintgausssqam.tex
-24.79

%% file: img/heapinttruncgausssqam.tex
-24.71

%% file: img/heapinthannsqam.tex
-23.14

%% file: img/heapinthammingsqam.tex
-22.66